\documentclass{amsart}
\usepackage{epsfig}
\usepackage{hyperref}
\usepackage{verbatim,amssymb,amsfonts,amsthm,amsmath,amscd,ifthen}

\def\sgn{\qopname \relax o{sgn}}

\def\pint{{- \kern-1.1em\int}}
\def\pist#1#2{\noindent\hangindent 2em\hangafter1\hbox to 2em{#1\hfil~~}#2}

\def\sqr#1#2{{\vcenter{\vbox{\hrule height .#2pt
                             \hbox{\vrule width .#2pt height#1pt \kern#1pt
                                   \vrule width .#2pt}
                             \hrule height .#2pt}}}}

{     \theoremstyle{plain}
        \newtheorem{theorem}{Theorem}[section]
         \newtheorem{lemma}[theorem]{Lemma} %[section]
}
{       \theoremstyle{remark}
        \newtheorem{remark}{Remark}[section]
}
\makeatletter
\@addtoreset{equation}{section}
\makeatother
\begin{document}

\renewcommand{\thetheorem}{\arabic{theorem}}
%{\arabic{section}.\arabic{theorem}}
\renewcommand{\theremark}{\arabic{remark}}
%{\arabic{section}.\arabic{remark}}
\renewcommand{\theequation}{\arabic{equation}} 
%{\arabic{section}.\arabic{equation}}
%
\def\too{{\dot{\to}}}
\def\trace{\mathop{\rm Trace}}
\def\N{\mathbb{N}}
\def\R{\mathbb{R}}
\def\E{\mathbb{E}}
\def\ls{\left[}
\def\ZP{{\mathbb{Z}}_{\geq0}}
\def\Z{{\mathbb{Z}}}
\def\C{{\mathbb{C}}}
\def\rs{\right]}\def\lp{\left(}\def\rp{\right)}
\def\diag{\qopname \relax o{diag}}
\def\vol{\qopname \relax o{vol}}
\def\Tr{\qopname \relax o{Tr}}
\def\tr{\qopname \relax o{tr}}
\def\dist{\qopname \relax o{dist}}
\def\dom{\qopname \relax o{Dom}}
\def\Prob{\qopname \relax o{Prob}}
\def\spec{\qopname \relax o{spec}}
\def\Ai{\qopname \relax o{Ai}}
\def\Aip{\Ai^\prime}
\def\xin{\xi^{(N)}}
\def\etan{\eta^{(N)}}
\def\ra{\rightarrow}
\def\eps{\epsilon}
\def\sgn{\qopname \relax o{sgn}}
\def\KA{K_{{\textrm{Airy}}}}
\def\const{{\rm const}\relax }
\def\sno{S_N^{(1)}}
\def\snop{S_N^{(1)\prime}}
\def\snf{S_N^{(4)}}
\def\snt{S_N^{(2)}}
\def\ph{\phi}
\def\corrf{\qopname \relax o{Corr}_N^{(4)}}
\def\corro{\qopname \relax o{Corr}_N^{(1)}}
\newcommand{\nc}{\newcommand}
\nc{\onetwo}[2]{\left(\begin{array}{cc}#1&#2\end{array}\right)}
\nc{\twotwo}[4]{\left(\begin{array}{cc}#1&#2\\&\\#3&#4\end{array}\right)}
\nc{\twoone}[2]{\left(\begin{array}{c}#1\\\\#2\end{array}\right)}
\nc{\mtt}[4]{\begin{pmatrix}#1&#2\\#3&#4\end{pmatrix}}
\newcommand{\bp}{\begin{proof}}
\newcommand{\ep}{\end{proof}}
\newcommand{\bt}{\begin{theorem}}
\newcommand{\et}{\end{theorem}}
\newcommand{\be}{\begin{equation}}
\newcommand{\ee}{\end{equation}}
\newcommand{\bq}{\begin{equation}}
\newcommand{\eq}{\end{equation}}
\newcommand{\ba}{\begin{aligned}}
\newcommand{\ea}{\end{aligned}}
\newcommand{\la}[1]{\label{#1}}
\nc{\fh}{\hat{f}}
\nc{\Fh}{\hat{F}}
\nc{\ft}{\tilde{f}}
\nc{\fth}{\hat{\tilde{f}}}
\nc{\pht}{\tilde{\phi}}
\nc{\pth}{\hat{\tilde{\phi}}}
\nc{\cd}{\cdots}
\nc{\ov}{\over}
\newcommand{\er}{\eqref}

\title[On the proof of universality]
{On the proof of universality for
orthogonal and symplectic ensembles
 in random matrix theory}

\author[Costin, Deift and Gioev]
{Ovidiu Costin, Percy Deift and Dimitri Gioev}
\address{Costin:
Department of Mathematics,
           The Ohio State University, 
           231 W. 18th Ave., 
           Columbus, OH 43210}
\email{costin@math.ohio-state.edu}
% Department of Mathematics, Rutgers University, 
%Piscataway, NJ 08854}%-8019}
%\email{costin@math.rutgers.edu}
%
\address{Deift: Department of Mathematics, Courant Institute of Mathematical
Sciences, New York University, 251 Mercer St., New York, NY 10012}
\email{deift@cims.nyu.edu}
\address{Gioev:
Department of Mathematics, University of Rochester, Hylan Bldg.,
Rochester, NY 14627}
\email{gioev@math.rochester.edu}
\begin{abstract}
We give a streamlined proof of a quantitative version
of a result from \cite{DG} which is crucial for the proof of
universality in the bulk \cite{DG}
and also at the edge \cite{DG2}
for orthogonal and symplectic ensembles of random matrices.
As a byproduct, this result gives asymptotic information on a certain ratio
of the $\beta=1,2,4$ partition functions for log gases.
\end{abstract}
\maketitle
For $m\geq2$, let
\be\la{2.1}
\ba
  h(x)&=\sum_{k=0}^{m-1}\beta_kx^{2k}\\
  \beta_k&=2\frac{(2m)(2m-2)\cdots(2m-2k)}
      {(2m-1)(2m-3)\cdots(2m-2k-1)},\qquad 0\leq k\leq m-1.
\ea
\ee
For odd $q$ set 
\begin{equation}
  \label{2.2}
  I(q)\equiv\frac{2}{\pi}\sin\frac{q\pi}{2}\int_{-1}^1\frac{\cos(q\arcsin
  x)}{h(x)(1-x^2)}\,dx
   =\frac{4}{\pi}\int_0^{\pi/2}\frac{\sin qs }{\sin s\,\, h(\cos s)}ds 
\end{equation}
and
\be\la{2.3}
   Q(q)\equiv I(q) + \frac1{2m}.
\ee
For $n\equiv 2m-1$, define the $(m-1)\times(m-1)$ matrix
\begin{equation}
  \label{eq:matr1}
  T^{[m-1]}= I - \frac{(m!)^2}{m(2m)!}Q^{[m-1]}B^{[m-1]}
            \equiv I - K^{[m-1]}
\end{equation}
where 
$$
  Q^{[m-1]}_{ij}=Q(n-2i+2j),\qquad
  B^{[m-1]}_{ij}=2m\binom{n}{j-i},\qquad 1\leq i,j\leq m-1.
$$
Here $\binom{n}{k}\equiv0$ for $k<0$.

In \cite[Theorem 2.6]{DG}, the authors prove the following result:
for $m\geq2$, 
\be\la{3.1}
      \det T^{[m-1]} \neq 0
\ee
(see Remark \ref{rem20.1} after the proof of Theorem \ref{thm1}
below). 
Note that in the notation of \cite{DG}, $T^{[m-1]}=T_{m-1}$.

In this paper we will give a streamlined proof
of the following quantitative version of \eqref{3.1}.
\begin{theorem}\la{thm1}
For $m\geq2$,
\be\la{3.2}
   \det T^{[m-1]} \geq 0.0865.
\ee
\end{theorem}
Equation \eqref{3.1} plays a crucial role in proving universality
in the bulk \cite{DG}, and also at the edge \cite{DG2},
for orthogonal ($\beta=1$) and symplectic ($\beta=4$)
random matrix ensembles for a class of weights
$w(x)=e^{-V(x)}$ where $V(x)$ is a polynomial 
$V(x)=\kappa_{2m}x^{2m}+\cdots$, $\kappa_{2m}>0$.
(Here $m$ is the same integer as in \eqref{3.1}, \eqref{3.2}.)
The situation is as follows.
In \cite{DG,DG2}, and also in \cite{DGKV},
the authors use the method of Widom \cite{W},
which is based in turn on \cite{TW2},
together with the asymptotic analysis for orthogonal
polynomials in \cite{DKMVZ2}.
A new and challenging feature of the method in \cite{W},
which does not arise in the proof of universality in the case $\beta=2$,
is the appearance of the \textit{inverse\ }of a certain matrix $C_{11}$
of fixed size $n=2m-1$ (see \cite[(1.37) and Theorem 2.3 et seq.]{DG}). 
In the scaling limit as $N\to\infty$,
the matrix $C_{11}$ converges to a matrix $C_{11}^\infty$
and
\be\la{4.1}
        \det C_{11}^\infty = (\det T^{[m-1]})^2
\ee
(see discussion from (2.13) up to Theorem 2.4 in \cite{DG}).
Thus in order to control the scaling limit for $\beta=1$ and $4$,
we need to show that $\det T^{[m-1]}\neq0$.

It turns out that $\det T^{[m-1]}$ is related to partition functions
for finite log gases in an external field $V$ at inverse temperatures
$\beta=1,2,4$
\be\la{5.1}
\ba
   Z_{V,\beta,k} &\equiv \frac1{k!} \int\cdots\!\int
                \prod_{1\leq i<j\leq k} |x_i-x_j|^\beta
                 e^{-\sum_{i=1}^k V(x_i)} \,dx_1\cdots dx_k\\
      &=\frac1{k!} \int\cdots\!\int
                e^{-\beta\sum_{1\leq i<j\leq k} \log|x_i-x_j|
                    -\sum_{i=1}^k V(x_i)} \,dx_1\cdots dx_k.
\ea
\ee
Using standard formulae for such partition functions
(see e.g.~\cite[(4.4), (4.17), (4.20)]{AvM}),
together with \cite[(2.18)]{DG}, one finds (see \cite[Remark 2.4]{St2},
\cite[Remark 1.5]{DG}) that for ensembles of (even) size $N$
\be\la{5.2}
        \det C_{11} = \bigg(\frac1{2^N(N/2)!}
      \frac{Z_{2V,4,N/2}\,Z_{V,1,N}}{Z_{2V,2,N}}\bigg)^2.
\ee
Thus
\be\la{6.1}
   \lim_{N\to\infty}
      \frac{Z_{2V,4,N/2}\,Z_{V,1,N}}{2^N(N/2)!\,Z_{2V,2,N}}
   =\det T^{[m-1]}\neq 0.
\ee
Formula \eqref{5.2}, together with \eqref{4.1},
 raises the possibility of using the methods of statistical
mechanics to prove \eqref{3.1}, \eqref{3.2}.
The estimates in \cite{J} show that the partition functions
$Z_{V,\beta,k}$ have, for certain constants $\alpha_{V,\beta}$,
 leading order asymptotics of the form
$e^{\alpha_{V,\beta}k^2(1+o(1))}$ as $k\to\infty$,
and moreover, their combined contributions to $\det C_{11}$
cancel to this order. In order to achieve cancellation
at subsequent orders, and so prove \eqref{3.1}, \eqref{3.2},
 one needs higher order asymptotics
for the $Z_{V,\beta,k}$'s, but, unfortunately
such asymptotics are
known only for $\beta=2$ (see \cite{EM}).
Regarding \eqref{5.2},
 we take the contrary point of view, i.e.,
 \eqref{6.1} and \eqref{3.2} provide new
information on the asymptotics of partition functions for log gases
at inverse temperatures $\beta=1$ and $4$.

Much of the analysis in \cite{DG} involves estimating $Q(q)$
in two regions: $3\leq q\lesssim \sqrt{m}$
and $\sqrt{m}\lesssim q\leq 4m-5$.
In this note, using bounds on
\be\la{7.1}
   W(x) \equiv \frac2\pi \int_0^x \frac{\sin qs}{\sin s}\,ds
\ee
which are uniform in $q=3,5,\cdots$ and in $0\leq x\leq \pi/2$
(see Lemma \ref{lem4} below), we are able to estimate $Q(q)$
uniformly for $q=3,5,\cdots,4m-5$ and so avoid
many of the technicalities in the proof in \cite{DG} of \eqref{3.1}.
Of course the function $W(x)$ is familiar from the analysis
of the Gibbs phenomenon in Fourier analysis.
\begin{remark}
For $m=1$, corresponding to the Gaussian orthogonal and
symplectic ensembles, $T^{[m-1]}$ is not defined
and no analog of \eqref{3.1}, \eqref{3.2} is needed (see \cite{DG}).
\end{remark}
\centerline{-----------------------}

We use the following result. For a matrix $X$
 let $r(X)=\sup\{|\lambda|:\lambda\in\spec X \}$
denote the spectral radius of $X$.
As is well \-known, for any operator norm $\|\cdot\|$ on $\{X\}$,
\be\la{8.0}
   r(X) = \lim_{j\to\infty} \|X^j\|^{1/j}= \inf_{j\geq1} \|X^j\|^{1/j}.
\ee 
\begin{lemma}
\la{lem2}
Assume $K$ and $K^\prime$ are $J$-dimensional matrices with
real entries such that $ |K_{ij}|\leq K_{ij}^\prime$,
$1\leq i,j\leq J$, and $r(K^\prime)<1$.
Then $r(K)<1$ and 
\be\la{8.1}
   \det(I-K)\geq \det(I-K^\prime)>0.
\ee
\end{lemma}
\begin{proof}
The following is true: if $r(X)<1$, then
\be\la{8p.1}
    \det(I-X) = e^{-\sum_{l=1}^\infty \frac1l\tr(X^l)}.
\ee
This result is usually stated in the form that \eqref{8p.1} holds
if $\|X\|<1$ (see e.g.~\cite{ReSi}).
To obtain \eqref{8p.1} for $r(X)<1$ from the case $\|X\|<1$
simply apply \eqref{8p.1} to $\mu X$ for $\mu$ small
and observe that for any fixed $\rho$ satisfying $r(X)<\rho<1$,
$\|X^l\|\leq \rho^l$ for $l$ sufficiently large: then
\eqref{8p.1} follows for $r(X)<1$ by analytic continuation $\mu\to1$.

Equip $\R^J$ with the $l_\infty$-norm $\|\cdot\|_\infty$ (any $l_p$-norm,
$1\leq p\leq \infty$ would do)
and for a matrix $X$ mapping $\R^J\to\R^J$
denote the associated operator norm by $\|X\|$.
For $\phi=\{\phi_j\}\in\R^J$ we denote the vector with
coordinates $\{|\phi_j|\}$ by $|\phi|$.
We claim that $r(K)\leq r(K^\prime)$. 
Indeed, for $\phi\in\R^J$, $|(K^l\phi)_j| \leq((K^\prime)^l|\phi|)_j$
and so 
$$ 
   \|K^l\phi\|_\infty \leq\|\, (K^\prime)^l|\phi|\, \|_\infty
        \leq\|(K^\prime)^l\| \,\, \| \,|\phi|\,\|_\infty
        =\|(K^\prime)^l\| \, \| \phi \|_\infty.
$$
Thus $\|K^l\| \leq\|(K^\prime)^l\|$ and so
$r(K)\leq r(K^\prime)<1$ by \eqref{8.0}.
It follows that \eqref{8p.1} is valid for $K$ and $K^\prime$.
But clearly $|\tr(K^l)| \leq\tr((K^\prime)^l)$
 and \eqref{8.1} is now immediate.
\end{proof}

The function $h(x)$ in \eqref{2.1} has the following properties
(see \cite[Proposition 6.2]{DG}): for $0<x<1$
\be\la{9.1}
\ba
   &(i)\quad h\textrm{ solves the differential equation}\\
   &\quad\qquad  x(x^2-1)h'+(2m-1-2(m-1)x^2 )h =4m\\
   &(ii)\quad \frac{4m}{2m-1}=h(0)\leq h(x)\leq h(1) =4m\\
    &(iii)\quad h(x)=\frac{4m x^{2m-1}}{\sqrt{1-x^2}}
              \int_x^1\frac{t^{-2m}}{\sqrt{1-t^2}}\,dt.
\ea
\ee
Property (i) reflects the fact that $h$ is a hypergeometric function,
$$
      h(x) = \frac{4m}{2m-1}{\space_2F_{1}}(1,-m+1,-m+3/2;x^2)
$$
(see \cite[(6.11)]{DG}) and (iii) follows by integrating (i).
Property (ii) follows from (i) and \eqref{2.1}.

Set 
\be\la{9.2}
       u(x)\equiv u(x;m)=\frac{1}{h(x)}-\frac{1-x^2}{2}+\frac{1}{4m}.
\ee
Note that the function $u(x)$ is closely related to the function
$y_m$ which plays a prominent role in \cite{DG}: we have
$$
   u(x) = \frac{\sqrt{1-x^2}}{m}\,y_m(\arcsin x) +
              \frac1{2m},\qquad 0\leq x\leq 1.
$$
Also note that using the elementary identities for $q=3,5,\cdots$,
$$
   \frac2\pi\int_0^{\pi/2} \sin qs\, \sin s\,ds = 0,\qquad
         W\Big(\frac\pi2\Big) = \frac2\pi\int_0^{\pi/2} 
            \frac{\sin qs}{ \sin s}\,ds = 1
$$
we have from \eqref{2.2}, \eqref{9.2}
\be\la{10.1}
   I(q) = \frac4\pi\int_0^{\pi/2} 
            \frac{\sin qs}{ \sin s}\,u(\cos s)\,ds - \frac1{2m}.
\ee
The main technical result in our proof of Theorem \ref{thm1}
is the following.
\begin{lemma}\la{lem3}
The function $u(x)=u(x;m)$, $m\geq2$, has the following properties.

(i) $u(x)$ is unimodal for $x\in[0,1]$. More precisely, there exists $x_0\in(0,1)$
such that $u^\prime(x)<0$ for $0<x<x_0$
and $u^\prime(x)>0$ for $x_0<x<1$.

(ii) $u(0)=0$, $u(1)=\frac1{2m}$.

(iii) For $0\leq x\leq 1$, 
$$
   -\frac1{4m}< u(x) \leq \frac1{2m}.
$$
\end{lemma}
The proof of Lemma \ref{lem3}
is given after the proof of Theorem \ref{thm1} below.
We also need the following elementary result from Fourier analysis.
\begin{lemma}
\la{lem4}
For $q\geq3$, $0\leq x\leq\pi/2$,
$$
       0\leq W(x)\leq \frac{\sqrt{3}}\pi + \frac23 < 1.218.
$$
\end{lemma}
\begin{proof}
As the factor $\sin s$ in $W(x)=\frac2\pi\int_0^x\frac{\sin qs}{\sin s}\,ds$
is increasing, a standard argument in the analysis of the Gibbs phenomenon
shows that for $0\leq x\leq\pi/2$,
$0\leq W(x)\leq\frac2\pi\int_0^{\pi/q}\frac{\sin qs}{\sin s}\,ds
=\frac2\pi\int_0^{\pi}\frac{\sin t}{q\sin(t/q)}\,dt$.
But for $0\leq t\leq\pi/2$, $q\mapsto q\sin(t/q)$ is increasing,
and so for $q\geq3$ and $0\leq x\leq\pi/2$,
 $0\leq W(x)\leq \frac2\pi\int_0^{\pi}\frac{\sin t}{3\sin(t/3)}\,dt
=\frac{\sqrt{3}}\pi + \frac23$.
\end{proof}
Assuming Lemma \ref{lem3}, we now prove Theorem \ref{thm1}.
By \eqref{2.3}, \eqref{10.1}, integrating by parts
 and using Lemma \ref{lem3}(ii),
$$
\ba
   Q(q) &= \frac{4}{\pi}
                        \int_0^{\pi/2}\frac{\sin qs }{\sin s}\, u(\cos s)\,ds \\
         &= 2 \big(W(s)\, u(\cos s)\big)\big|_0^{\pi/2} 
                            + 2\int_0^{\pi/2} W(s)\,  u^\prime(\cos s)\,\sin s\,ds\\
    &=  2\Big(\int_0^{\arccos x_0}
             +\int_{\arccos x_0}^{\pi/2} \Big) W(s)\,  u^\prime(\cos s)\,\sin s\,ds.
\ea
$$
Thus, by Lemma \ref{lem3} and Lemma \ref{lem4}, %for $m\geq16$,
$$
\ba
   Q(q) &\leq 2\int_0^{\arccos x_0} W(s)\,  u^\prime(\cos s)\,{\sin s}\,ds\\
            &\leq 2\cdot1.218 \cdot \big( u(1) - u( x_0)\big)\\
           & \leq 2\cdot1.218 \cdot \Big( \frac1{2m}+\frac1{4m}\Big)
           =\frac32\cdot \frac{ 1.218}m.
\ea
$$
On the other hand
$$
\ba
   Q(q) &\geq 2\int_{\arccos x_0}^{\pi/2} W(s)\,  u^\prime(\cos s)\,{\sin s}\,ds\\
            &\geq 2\cdot1.218 \cdot \big( u(x_0) - u(0)\big)
           \geq -\frac12\cdot \frac{ 1.218}m
\ea
$$
and thus 
$$
             |Q(q)|\leq \frac32\cdot \frac{ 1.218}m = \frac{ 1.827}m.
$$
Recalling the definitions of $Q^{[m-1]}$ and $B^{[m-1]}$,
we have for $1\leq i,j\leq m-1$
$$
\ba
    \big| (Q^{[m-1]} B^{[m-1]})_{ij}\big|
       &\leq \Big|\sum_{l=1}^j Q(n-2i+2l)\, 2m\binom{n}{j-l}\Big|
          \leq 2\cdot 1.827\cdot \sum_{l=0}^{j-1} \binom{n}{l}
\ea
$$
and hence
$$
\ba
    \big| (Q^{[m-1]} B^{[m-1]})_{ij}\big|
       \leq  2\cdot 1.827\cdot L_{ij},\qquad 1\leq i,j\leq m-1,
\ea
$$
where $L$ is the rank $1$ matrix with entries
   $L_{ij}=\sum_{l=0}^{j-1} \binom{n}{l}$, independent of $i$.
Hence $L$ has only $1$ non-zero eigenvalue $\lambda_1(L)$
and we find
\be\la{12.1}
\ba
    r(L) &= \lambda_1(L) = \sum_{k=1}^{m-1} L_{1k}
        = \sum_{k=1}^{m-1} \sum_{l=1}^{k} \binom{n}{l-1}\\
      &= \sum_{l=0}^{m-1} (m-l-1)\binom{2m-1}{l} 
   = \frac{m}2\binom{2m-1}{m-1} - 2^{2m-3}
       \leq \frac{m}2\binom{2m-1}{m-1}.
\ea
\ee
In the second last step, we have used the elementary formula
preceding (6.7) in \cite{DG}.

Assembling the above results and recalling the definition
of $K^{[m-1]}$, we obtain for $1\leq i,j\leq m-1$,
$$
    \big| K^{[m-1]}_{ij}\big|
       = \frac{(m!)^2}{m(2m)!}\,
                      \big| (Q^{[m-1]} B^{[m-1]})_{ij}\big|
       \leq  K_{ij}^\prime
$$
where 
$$
          K_{ij}^\prime \equiv 2\cdot 1.827\cdot\frac{(m!)^2}{m(2m)!}\,L_{ij},
            \qquad 1\leq i,j\leq m-1,
$$
and by \eqref{12.1}, the only non-zero eigenvalue of $K^\prime$
satisfies
\be\la{13.1}
\ba
     \lambda_1(K^\prime) &= r(K^\prime)
          = 2\cdot 1.827\,\frac{(m!)^2}{m(2m)!}\,r(L)\\
     &\leq 2\cdot 1.827\,\frac{(m!)^2}{m(2m)!}\,
                          \frac{m}2\binom{2m-1}{m-1}
%\\   &
=\frac{1.827}2 = 0.9135 < 1.
\ea
\ee
Thus by Lemma \ref{lem2}, 
$$
    \det(1-K^{[m-1]}) \geq \det(1-K^{\prime}) = 1-\lambda_1(K^\prime)
                    \geq 0.0865.
$$
This completes the proof of Theorem \ref{thm1}. % for $m\geq16$.
\begin{remark}\la{rem20.1}
Using Lemma \ref{lem2}, the calculations in \cite{DG}
also yield a quantitative version of \eqref{3.1}
but with a weaker bound.
As above, we estimate $T^{[m-1]}$ elementwise
with a {\em rank one\ }matrix
so that we can estimate the
determinant by estimating the only nonzero eigenvalue.
We note that we cannot use \cite[(6.22)]{DG}
 (the matrix in (6.22) is not rank one).
For ``small'' $m$
we use \cite[(6.15), (6.16)]{DG}, %as in \eqref{eq_aux} above,
and for ``large'' $m$ we use
\cite[(6.55), (6.21)]{DG}. 
We claim that 
\be\la{DG_conseq}
   \det T^{[m-1]} \geq 0.02,\qquad m\geq 2.
\ee
This estimate is not optimal, but we could not strengthen it compared
 to \eqref{3.2} by the methods in \cite{DG}.
To prove \eqref{DG_conseq} for $2\leq m\leq 46$,
we note that the RHS in \cite[(6.16)]{DG} is $<0.98$ for $m$ in this range.
(Note that our $Q(q)$ and $\tilde{I}(q)$ in \cite{DG}
are related by $\tilde{I}(q) = mQ(q) - 1$
and hence $|Q(q)|=\frac1m|1+\tilde{I}(q)|\leq\frac1m(1+|\tilde{I}(q)|)$.)
To prove \eqref{DG_conseq} for $m\geq 47$, we 
set $\delta\equiv 0.04$ and consider $q=3,5,\cdots 4m-5$
in the regions $\frac4\pi\frac{\sqrt{m+1/2}}{q}\leq1-\delta$
and $\frac4\pi\frac{\sqrt{m+1/2}}{q}>1-\delta$ separately. 
In the former
$q$-region, by \cite[(6.21)]{DG}, 
$
        |1+\tilde{I}(q)| \leq 1+|\tilde{I}(q)| \leq 1.96.
$
In the latter $q$-region, substituting
 $\frac{q}{\sqrt{m+1/2}}\leq \frac4\pi\frac1{1-\delta}$
in \cite[(6.55)]{DG}, we note that the resulting estimate on 
$|1+\tilde{I}(q)|$ multiplied by
 $(\frac12 - \frac{(m!)^2}{m(2m)!}\,2^{2m-2})$,
is $<0.98$ in fact for $m\geq44$.
These facts together with Lemma \ref{lem2}
prove \eqref{DG_conseq} 
%(cf.~\eqref{eq_aux} and \cite[(6.56), (6.57)]{DG}).
(cf.~\cite[(6.56), (6.57)]{DG}).
\end{remark}
It remains to prove Lemma \ref{lem3}.
A straightforward computation using \eqref{9.1}(i)
 and \eqref{9.2} shows that $u$ is a solution of the equation
\be\la{14.1}
   x(1-x^2)u^\prime - 4mu^2 + (2(m+1)x^2 + 1-2m) u - \frac{x^2}{2m} = 0.
\ee
Moreover as $h(x)>0$, $u$ is smooth.
By \eqref{9.1}(ii), and by differentiating \eqref{14.1}, we find,
\be\la{14.2}
\ba
   &u(0)=0,\qquad u^\prime(0) = 0, \qquad 
           u^{\prime\prime}(0) = -\frac1{m(2m-3)}\\
   &u(1) = \frac1{2m},\qquad u^\prime(1) = \frac23+\frac1{3m}.
\ea    
\ee
Now observe that at a point $0<x<1$ where $u^\prime(x)=0$,
we cannot have $4m(m+1)u(x)-1=0$,
i.e.~$u(x)=\frac1{4m(m+1)}$. Indeed, substituting
these values into \eqref{14.1}, 
we find $-1+(1-2m)(m+1)=0$, which is a contradiction.
Next we show that
\be\la{15.1}
   \big(u^{\prime}(x)=0\textrm{ for some }0<x<1\big)
   \quad\implies\quad u^{\prime\prime}(x)
      = \frac{(4m(m+1)u(x)-1)^2}{m(1-2mu(x))(1-4mu(x))}.
\ee
Indeed, differentiating \eqref{14.1}, we find for such a point $x$
\be\la{15.2}
     u^{\prime\prime}(x)
      = \frac{1-4m(m+1)u(x)}{m(1-x^2)}.
\ee
Setting $u^\prime(x)=0$ in \eqref{14.1} and solving for $(1-x^2)$
in terms of $u(x)$, we obtain
\be\la{15.3}
      1-x^2 = - \frac{(1-2mu(x))(1-4mu(x))}{4m(m+1)u(x)-1}.
\ee
Note that by the above argument,
the denominator in \eqref{15.3} is non-zero: also the numerator is non-zero
as $1-x^2\neq0$. Substituting \eqref{15.3} into \eqref{15.2}
we obtain \eqref{15.1}.
Furthermore, the calculation shows that if $u^\prime(x)=0$
for some $0<x<1$, then $u^{\prime\prime}(x)$ is (finite and) non-zero.

From \eqref{14.2} we see that for small $x>0$, $u(x)<0$.
As $u(1)>0$, there must be at least one point $x\in(0,1)$ where $u(x)=0$.
But it follows from \eqref{14.1} that if $u(x)=0$, $x\in(0,1)$, then
$u^\prime(x)=\frac{x}{2m(1-x^2)}>0$. Hence $u$ crosses the level zero
at a unique point $x_1\in (0,1)$.
Next suppose that $u^\prime(\hat{x})=0$ for some $\hat{x}\in(0,x_1)$.
But then
by \eqref{15.1}, $u^{\prime\prime}(\hat{x})>0$ as $u(\hat{x})<0$.
Thus any critical point for $u(x)$ in $(0,x_1)$, must be a local minimum.
As $u(x)$ clearly has a minimum on $(0,x_1)$,
it follows that it has a unique minimum at $x_0\in(0,x_1)$, say,
and no other critical points on $(0,x_1)$.
Thus $u^\prime(x)<0$ for $0<x<x_0$, and
$u^\prime(x)>0$ for $x_0<x\leq x_1$.

Next we show that
\be\la{16.1}
                  0<u(x)<\frac1{2m}\qquad \textrm{for } x_1<x<1.
\ee
Indeed, if $u(x)=\frac1{2m}$ for $0<x<1$, then from \eqref{14.1}
we find $u^\prime(x)=\frac{2m+1}{2mx}>0$.
But we know from \eqref{14.2} that $u(1)=\frac1{2m}$, $u^\prime(1)>0$.
Hence $u(x)$ cannot cross the level $\frac1{2m}$ for $0<x<1$.
This proves \eqref{16.1}.

To complete the proof that $u$ is unimodal we show that
$u^\prime(x)>0$ for $x_1<x<1$.
Suppose $u^\prime(x_2)<0$ for some $x_1<x_2<1$.
Then as $u(x_1)=0$ and $u(x_2)<u(1)=\frac1{2m}$, there
must exist $x_1<x_2^-<x_2$ and $x_2<x_2^+<1$
such that $u$ has a local maximum at $x_2^-$
and a local minimum at $x_2^+$.
By \eqref{15.1}, we must have $u(x_2^-)>\frac1{4m}$
and $u(x_2^+)<\frac1{4m}$.
This implies, in particular, that $u(x)$ crosses the level $\frac1{4m}$
at at least one point $x^{\#}\in(x_2^-,x_2^+)$ such that
 $u^\prime(x^\#)\leq0$. 
But by \eqref{14.1}, $u(x)=\frac1{4m}$, $0<x<1$, implies
$u^\prime(x)=\frac1{2x}>0$, which is a contradiction.
Thus $u^\prime(x)\geq0$ on $(x_1,1)$.
On the other hand if $u^\prime(x_3)=0$ for some $x_1<x_3<1$,
then by \eqref{15.1}, $u^{\prime\prime}(x_3)\neq0$ and so
$u^\prime(x)$ changes sign in a neighborhood of $x_3$,
contradicting $u^\prime(x)\geq0$ on $(x_1,1)$.
Thus $u^\prime(x)>0$ for all $x_1\leq x\leq 1$.
This completes, in particular, the proof of part (i) of Lemma \ref{lem3}.

It remains to show that $u(x)=u(x;m)>-\frac1{4m}$
for $m\geq2$, $x\in[0,1]$.
It turns out that $x=x_m\equiv\sqrt{\frac{m-1}{m+2}}$ plays a 
distinguished role. More precisely, as we now show,
\be\la{eq_star}
  u(x_m)>-\frac1{4m}
   \quad\implies\quad\Big( u(x)>-\frac1{4m}\textrm{ for all }x\in[0,1]\Big).
\ee
To see this, suppose $u(x)=-\frac1{4m}$
for some $x\in(0,1)$: then from \eqref{14.1} we obtain
\be\la{eq_stst}
   u^\prime(x) = \frac{(m+2)x^2 - (m-1)}{2mx(1-x^2)}.
\ee
Suppose $u(x_m)>-\frac1{4m}$.
 If $u(\hat{x})\leq-\frac1{4m}$
for some $0<\hat{x}<x_m$, then clearly $u(x^{\#})=-\frac1{4m}$,
$u^\prime(x^{\#})\geq0$ for some point
$x^{\#}\in[\hat{x},x_m)$. But by \eqref{eq_stst}, $u^\prime(x^{\#})<0$,
which is a contradiction.
 Similarly if $u(\hat{x})\leq-\frac1{4m}$
for some $x_m<\hat{x}<1$, there must exist 
a point $x^{\#}\in(x_m,\hat{x}]$
such that $u(x^{\#})=-\frac1{4m}$,
$u^\prime(x^{\#})\leq0$. But this contradicts \eqref{eq_stst}
as above. This proves \eqref{eq_star}.

To complete the proof of Lemma \ref{lem3}, we must prove 
$u(x_m)\equiv u(x_m;m)>-\frac1{4m}$, $m\geq2$.
Set $s=1-x$. From \eqref{9.1}(iii), we obtain
$$
\ba
    h(x)  = \frac{4m(1-s)^{2m-1}}{\sqrt{s(2-s)}}
                          \int_0^s \frac{(1-\tau)^{-2m}}{\sqrt{\tau(2-\tau)}}\,d\tau%\\
      \leq \frac{4m(1-s)^{2m}}{(2-s)(1-s)\sqrt{s}}
                          \int_0^s \frac{(1-\tau)^{-2m}}{\sqrt{\tau}}\,d\tau.
\ea
$$
Using the elementary inequality $\frac{1-s}{1-\tau}\leq e^{\tau-s}$
for $0\leq \tau\leq s\leq1$, we find
$$%\be\la{18.3}
    h(x) \leq  \frac{4me^{-2ms}}{(2-s)(1-s)\sqrt{s}}
                          \int_0^s \frac{e^{2m\tau}}{\sqrt{\tau}}\,d\tau
               = \frac{4me^{-\mu^2}}
                                  {(1-\frac{\mu^2}{4m})(1-\frac{\mu^2}{2m})\mu}
                           \int_0^\mu e^{\lambda^2}\,d\lambda
$$%\ee
where 
$$%\be\la{18.4}
                 \mu = \sqrt{2ms} = \sqrt{2m(1-x)}.
$$%\ee

In order to prove $u(x_m)>-\frac1{4m}$, $m\geq2$,
 we see that it is sufficient to show that
$$%\be\la{19.1}
    \frac{(1-\frac{\mu^2}{2m})\mu e^{\mu^2}}
                 { 2\int_0^\mu e^{\lambda^2}\,d\lambda}
              -\mu^2 + \frac1{1-\frac{\mu^2}{4m}} >0\qquad
           \textrm{for }\mu=\mu_m=\sqrt{2m(1-x_m)}.
$$%\ee
By the inequality $(1-\frac{\mu^2}{4m})^{-1}>1+\frac{\mu^2}{4m}$,
and the elementary fact that $1<\mu_m<\sqrt{3}$, $m\geq2$,
we see that it is sufficient to show
$$%\be\la{19.2}
   F(\mu) \geq \frac1{m}\,G(\mu)\qquad
           \textrm{for }1\leq\mu\leq\sqrt{3}
$$%\ee
where 
$$%\be\la{19.3}
\ba
   F(\mu)\equiv \mu e^{\mu^2}
               +2(1-\mu^2) \int_0^\mu e^{\lambda^2}\,d\lambda,\qquad%\\
   G(\mu)\equiv \frac{\mu^2}2
               \Big(  \mu e^{\mu^2} - \int_0^\mu e^{\lambda^2}\,d\lambda\Big).
\ea
$$%\ee
But $G(\mu)$ is clearly increasing and so it is enough to show
\be\la{19.4}
   F(\mu) \geq \frac{G(\sqrt{3})}{m}\qquad
           \textrm{for }1\leq\mu\leq\sqrt{3}.
\ee
Differentiating $F(\mu)$ we find
$$
\ba
   F(1) &= e,\qquad 
   F^\prime(1) = 3e-4 \int_0^1 e^{\lambda^2}\,d\lambda > 2.304 >0\\
   F^{\prime\prime}(1) &= 2e-4 \int_0^1 e^{\lambda^2}\,d\lambda > -0.415\\
   F^{\prime\prime\prime}(\mu) &\geq 0\qquad\textrm{for }\mu\geq1.
\ea   
$$
Thus for $1\leq\mu\leq\sqrt{3}$
$$
         F(\mu) \geq F(1) + F^\prime(1)(\mu-1)
                   +\frac{F^{\prime\prime}(1)}{2}(\mu-1)^2
        \geq e - \frac{0.415}{2}(\sqrt{3}-1)^2 > 2.607.
$$
On the other hand $G(\sqrt{3})<41.3$, and if we choose $m$ so that
$2.607>\frac{41.3}{m}$, then \eqref{19.4} will hold.
Clearly $m\geq16$ satisfies this inequality.
We conclude that $u(x_m)>-\frac1{4m}$ for $m\geq16$.
On the other hand, using Maple (only sums and products are
involved),
 we find from \eqref{2.1}, \eqref{9.2}
$$
    \min_{2\leq m\leq15} \Big(u(x_m)+\frac1{4m}\Big) > 0.0129 > 0.
$$
This completes the proof of Lemma \ref{lem3},
and hence Theorem \ref{thm1}.

{\bf Acknowledgments. }The authors would like to thank Thomas Kriecherbauer
for useful conversations.
The work of the 
first author was supported in part by NSF grants
DMS--0103807 and DMS--0100495. 
The work of the second author was supported in part by
NSF grants DMS--0296084 and 
DMS--0500923.
While this work was being completed, the second author
was a Taussky--Todd and Moore Distinguished Scholar at Caltech,
and he thanks Professor Tombrello for his sponsorship
and Professor Flach for his hospitality.
The work of the third author was supported in part by
the NSF grant DMS--0556049.
The third author would like to thank
the Courant Institute
and Caltech for hospitality
and financial support. 
Finally, the third author would like
to thank the Swedish foundation STINT 
for providing basic support to visit Caltech.

\end{document}